\documentclass[a4paper,12pt]{article}

\def\al   {\alpha}

\def\bet  {\beta}

\def\HH {\hat H}

\def\HS {\hat S}

\def\HH {\hat H}

{\it}
{\it}
{\it}
{\it}
{\it}
{\it}

\def\bet {\beta}

\def\de {\delta}

\usepackage{times}
\usepackage{graphicx}
\begin{document}
\title{ A simple method to compute excitation energies in variational 
 many-body calculations.}
\author{G. Puddu\\
       Dipartimento di Fisica dell'Universita' di Milano,\\
       Via Celoria 16, I-20133 Milano, Italy}
\maketitle
\begin {abstract}
         We propose a simple, easy to implement, variant of the 
         EXCITED method for variational many-body calculations for excited states.
         We apply this method to the Hybrid Multideterminant method(HMD). We test
         this method with relatively few Slater determinants by comparing the results 
         with exact  shell model calculations  for ${}^{56}Ni$ using the $fpd6$ interaction.   
         We obtain very good agreement with the exact results.
\par\noindent
{\bf{Pacs numbers}}: 21.60.Cs,$\,\,$  24.10.Cn,$\,\,$  27.40.+z
\vfill
\eject
\end{abstract}
\section{ Introduction.}
     In performing variational calculations for ground-state energies for many-body nuclear
     Hamiltonians, quite often we are left with the problem of obtaining reasonable
     values for the excitation energies. In the Hybrid Multideterminant method (HMD) (ref.[1]) for
     example, we assume that the ground-state wave function is well described by a sum of
     (usually projected to good angular momentum and parity)  generic (that is no restrictions
     are imposed) Slater determinants.
     Both the Slater determinants and the coefficients of the linear combination are determined
     by minimizing the expectation values of the Hamiltonian. By increasing the number 
     of the Slater determinants we obtain increasingly accurate values of the ground-state
     energy as compared to the exact ones. Equivalently, after the minimization procedure has 
     been carried out, we can diagonalize the many-body Hamiltonian in the basis of these
     Slater determinants and obtain a set of eigenvalues. However all eigenvalues, except the
     ground-state, give a rather poor description of the excitation energies, unless
     the basis consists of a very large number of Slater determinants. This is not very surprising
     since the basis has been determined so as to minimize the ground-state energy, and therefore
     all other eigenvalues are usually much higher than the exact ones. More than $20$ years ago, a method to 
     remove
     this deficiency has been proposed in variational calculations using quasi-particle determinants, that is,
     the EXCITED-MAD-VAMPIR approach (ref.[2]). The method consists in first to obtain the ground-state wave function,
     with the appropriate number of particles, angular momentum and parity, and then  to determine the first excited
     state imposing at every step orthogonality (with the Gram-Schmidt method) to the ground-state wave 
function.
     The method is iterated until all the desired excited state are obtained. We have applied this EXCITED
     method to the HMD method. Our version of the computer program  increases the corresponding ground-state
     by about 170 lines in FORTRAN 77.
     We found however a much simpler method to obtain excited states which can be implemented
     by the addition of very few lines to the computer program for the ground-state. Moreover if, after
     the calculation has been finished, we wish to increase the accuracy of the low energy wave
     functions, in the EXCITED method we would presumably have to repeat  the calculation. The method
     we propose does not have this inconvenience and it is very flexible and easy to implement.
     The method is based on the idea that in order to increase the sensitivity of the minimization method on the 
     position of the excited states energies(rather than just on the ground-state energy) the quantity to be minimized is
     some weighed average of the the first few low-lying eigenvalues obtained from the diagonalization
     of the Hamiltonian in the basis of the Slater determinants. For example for three states the functional
     to be minimized is $ F= (w_1E_1+w_2 E_2+w_3E_3)/(w_1+w_2+w_3)$ with the weights $w_i>0$.
     We call this method the  Centroid HMD method (CHMD).
     It takes very few lines to modify the ground-state computer program to implement this idea.
\par
      This paper is organized as follows. In section 2 we review in more detail the EXCITED method
     (see however the original work of ref.[2]) and the centroid method. In section 3
     we test the method by comparing the exact energies of ${}^{56}Ni$ recently obtained by shell model diagonalization
     in refs. [3],[4].
\bigskip
\bigskip
\section{ The EXCITED and the Centroid method.}
\bigskip
     In the EXCITED method, that we apply here to Slater determinants rather than to quasi-particle
     states, we assume to have determined an accurate approximation to the ground-state of the many-body Hamiltonian $\HH$ 
     as
$$
|\psi(1)> = |\phi(1)>=\sum_{\al=1}^{N_1}g_{\al,1}|U_{\al,1}>,
\eqno(1)
$$
     where the $|U_{\al,1}>$ are variational Slater determinants with good quantum numbers (angular momentum, 
etc.)
     restored with projectors. $N_1$ is the number of Slater determinants necessary for good convergence to the 
     ground-state energy (for non-zero angular momentum it includes also the $K$ quantum number). 
     The coefficients $g_{\al,1}$ are determined by solving the generalized eigenvalue problem
$$
\sum_{\bet} <U_{\al,1}|\HH|U_{\bet,1}> g_{\bet,1} = E_1 \sum_{\bet} <U_{\al,1}|U_{\bet,1}> g_{\bet,1},
\eqno(2)
$$
     $E_1$ being the lowest eigenvalue. All other eigenvalues obtained in eq.(2) are discarded 
     since they are a poor approximation to the energies of the excited states.
     The Slater determinants in eq.(1) can be generated with a two steps method.
     In the first step (addition step) Slater determinants are added to the basis one by one
     and each Slater determinant is varied so as to minimize the ground-state energy. In the second
     step (refinement step) once the basis is constructed, each Slater determinant is varied anew (ref.[5]).
     The motivation for this second step is that the first Slater determinants have been varied
     when the basis contained few Slater determinants. Ideally, we would want to minimize the energy
     by varying all Slater determinants simultaneously, while in the addition step only the last added
     one is varied. The refinement step corrects this partial variation.
\par   
     For the excited states one makes use of a variational ansatzs similar to eq.(1)
$$
|\phi(n)>= \sum_{\al=1}^{N_n}g_{\al,n}|U_{\al,n}> \;\;\;\;\;(n=2,..,N_n),
\eqno(3)
$$
     and the excited states are written as
$$
|\psi(n)>=\bet_{nn} \big  |\phi_n>+ \sum_{j=1}^{n-1} \bet_{jn} |\phi_(j)>,
\eqno(4)
$$
     the coefficients $\bet_{i,j}$ ($i,j=1,..,n$) are determined so that $<\psi_i|\psi_j>=\de_{i,j}$.
     The wave functions in eq. (3) are determined by minimizing
$$
E_n={<\psi(n)| \HH |\psi(n)> \over <\psi(n)|\psi(n)>}.
\eqno(5)
$$
     In the above equation only the $g_{\al,n}|U_{\al,n}>$ are varied, all previous $g_{\al,i}|U_{\al,i}>$ for
     $i=1,..,n-1$ are kept fixed and the coefficients $\bet_{in}$ ($i=1,..,n-1$) are determined by the orthogonality
     condition. The result of the Gram-Schmidt method can be recast in a compact form as follows. As in ref. [2], if we 
     define the overlap matrix between previously determined non-orthogonal wave functions as 
$$
A_{i,j}=<\phi(i)|\phi(j)>\;\;\;\;(i,j,=1,...,n-1),
\eqno(6)
$$
     and its inverse $B\equiv A^{-1}$ of dimension $n-1$ and the Gram-Schmidt projector as
$$
\HS=\sum_{i,j=1}^{n-1} \phi(i)>B_{ij} <\phi(j)|,
\eqno(7)
$$
     then the $n^{th}$ state given by the Gram-Schmidt procedure is
$$
|\psi(n)>=(1-\HS)|\phi(n)>,
\eqno(8)
$$
     and the quantity to be minimized is 
$$
 E_n={ <\phi(n)|(1-\HS)\HH (1-\HS)|\phi(n)>\over<\phi(n)|(1-\HS)|\phi(n)>},
\eqno(9)
$$
    that is
$$
E_n={H_{nn}-H_{ni} B_{ij}A_{jn}-A_{n,i}B_{ij}H_{j,n} +A_{ni}B_{i,j}H_{jl}B_{jk}A_{kn} \over
  A_{nn} -A_{ni}B_{ij}A_{jn} }.
\eqno(10)
$$
    In the above equation the sum over repeated indices, except $n$, is understood, and $H_{i,j}=<\phi(i)|\HH|\phi(j)>$.
    The quantities $\overline H_{ik}=B_{i,j}H_{jl}B_{jk}$ can be stored in the previous calculations.
    All overlaps and matrix elements in eq.(10) are angular momentum projected, and their evaluation
    is certainly the most expensive part of the calculation.
    The minimization step is performed with
    quasi-newtonian methods (cf. ref.[6]-[8] and references in there) once the derivatives of $E_n$ with
    respect to the entries of the Slater determinants have been determined.
    Clearly after the calculation of several levels has been performed if we wish to have more accurate wave-functions
    we would have to determine anew at least some of the eigenstates.
\par
    In the CHMD method instead we use only the ansatz
$$
|\phi> = \sum_{\al=1}^{N}g_{\al}|U_{\al}>,
\eqno(11)
$$
     and, after the matrix elements $h_{\al',\al}=<U(\al'|\HH|U_{\al}>$ and $o_{\al',\al}=<U(\al'|U_{\al}>$
     have been  evaluated we solve the generalized eigenvalue problem, written in a matrix notation as
$$
h g = o g E,
\eqno(12)
$$
     and obtain all $g(\al,k)$ for $\al,k=1,..,N$ and the energies $E_k$. The input to the minimization 
     procedure is then $F=\sum_k w_k E_k/\sum w_k$ with positive weights $w_k$. We typically take only few
     states in the sum. As a rule of the thumb $w_1>w_2 >w_3$ etc. especially for a small number of Slater
     determinants.  The reason is that the location of the excited states 
     is very sensitive to the values of $w_2,w_3,.$.
     If we take comparable values for the weights, we could have an increase
     (rather than a decrease) of the ground-state energy, even though the centroid would decrease, because of the large
     contribution of the higher eigenvalues of eq.(12). Therefore  
     we prefer  to assign a large weight to the ground-state so a decrease in the centroid generally 
     gives a decrease of all energies included in the sum. Only when the energies are sufficiently refined
     we can take comparable values for the weights.
\par
     There is a very interesting limit of this method, namely when all $w_k=0$ except one.
     In this case we first determine the ground-state with $w_1=1$ (using $N_1$ Slater determinants) then we keep 
     adding $N_2$ Slater determinants to the basis setting $w_1=0,w_2=1$ (the dimension 
     of eq. (12) is now $N_1+N_2$), then we proceed setting
     $w_1=w_2=0,w_3=1$ and add $N_3$ additional  Slater determinants. Of course we do not vary the Slater determinants
     which belong to eigenstates different from the desired one. In this way we simulate the EXCITED method but we
     can improve at wish previously determined eigenstates without any concern for orthogonality which is automatically
     taken care by the diagonalization of eq.(12).
     Note that if we generate eigenstates in sequence by adding Slater determinants one after the other, that is if we use
     only the addition step without the refinement step,
     we need no new computer program at all. In fact, after $N_1$ Slater determinants have been generated for the 
     ground-state, 
     we can simply select  the first excited state in eq. (12), the label of the desired eigenstate  being an 
     input 
     parameter and keep adding more Slater determinants to the basis, minimizing $E_2$ without any modification
     to the computer program.
\par
     Note that the number of projected matrix elements to be computed in this limit, is the same as in the EXCITED method.
     When we are varying the Slater determinants for the excited states, the ground-state energy 
     decreases although by small amounts. The main difference between this limiting case (that is all $w$'s set to $0$
     except one) and the EXCITED method is that we make use of eq.(12) in a basis consisting of all $N_1+N_2+N_3+..$
     Slater determinants, while the EXCITED method  uses only $N_2$ Slater determinants for the first excited state, $N_3$  
     for the second excited state and so on. In the next section we apply the CHMD method to the case of ${}^{56}Ni$.
\bigskip
\bigskip
\section{ ${}^{56}Ni$.}
\bigskip
     As an application of the CHMD method we consider ${}^{56}Ni$ with an inert ${}^{40}Ca$ core, the Hamiltonian
     for the valence nucleons is the $fpd6$ Hamiltonian (ref. [9]). This case has been investigated recently with shell model
     diagonalization in the full space in ref.[3]. The energies for the $0^+_1,0^+_2$ and $0^+_3$ states are
     $E(0^+_1)=-203.198$MeV, $E(0^+_2)=-198.723$MeV, $E(0^+_3)=-198.204$MeV (ref.[4]). 
     For the $2^+_1$ state the shell model value is $E(2^+_1)=-200.190$MeV.  
     Since we have only three states from the shell model calculation we consider a centroid with only the first three
     states. Let us consider first the limiting case where only one of the $w$'s is non-zero. We consider
     the following sequence. First we generate $15$ Slater determinants with $w_1=1,w_2=w_3=0$, and obtained
     $E(0^+_1)=-203.154$MeV, $E(0^+_2)=-197.632$MeV, $E(0^+_3)=-193.330$MeV. As discussed in the introduction,
     although the ground-state energy is well approximated (only $44$KeV above the exact value), the first excited
     $0^+$ state is about $1.1 $ MeV above the exact value and even worse the second excited $0^+$ state is about $4.9$ MeV 
     above  the exact value.
\par\noindent
     We then added $15$ more Slater determinants to the CHMD basis with $w_1=0,w_2=1,w_3=0$ and obtained 
     $E(0^+_1)=-203.157$MeV, $E(0^+_2)=-198.597$MeV, $E(0^+_3)=-197.881$MeV. The first excited $0^+$ state is
     now only $126 KeV$ above the exact value and also the energy of the $0^+_3$ state has dramatically improved.
     We added then  $15$ more Slater determinants to the CHMD basis with $w_1=w_2=0,w_3=1$ and obtained
     $E(0^+_1)=-203.160$MeV, $E(0^+_2)=-198.603$MeV, $E(0^+_3)=-198.147$MeV. The discrepancy for the $0^+_3$ state
     is now only $57$ KeV, and also the energies for the $0^+_1$ and $0^+_2$ states have improved.
     With only $45$ Slater determinants we obtained very good values of the energies for $3$ states.
     We decided then to improve on the ground-state and on the first excited $0^+$ state. The full sequence of
     calculation is now $15$ Slater determinants for $0^+_1$ , $15$ Slater determinants for $0^+_2$,
     $15$ Slater determinants for $0^+_3$,  $5$ more Slater determinants for the $0^+_1$ state and  
     $5$ more Slater determinants for  the $0^+_2$ state. The energies thus obtained are $E(0^+_1)=-203.166$MeV, 
     $E(0^+_2)=-198.630$MeV, 
     $E(0^+_3)=-198.132$MeV. The discrepancies from the exact values are  $32$KeV,$93$KeV and $72$KeV for
     the $0^+_1,0^+_2,0^+_3$ states respectively. In fig. 1 we show the behavior of the energies as a function of
     of the number of Slater determinants. The effect of the changes in the weights is clearly recognizable.
\renewcommand{\baselinestretch}{1}
\begin{figure}
\centering
\includegraphics[width=10.0cm,height=10.0cm,angle=0]{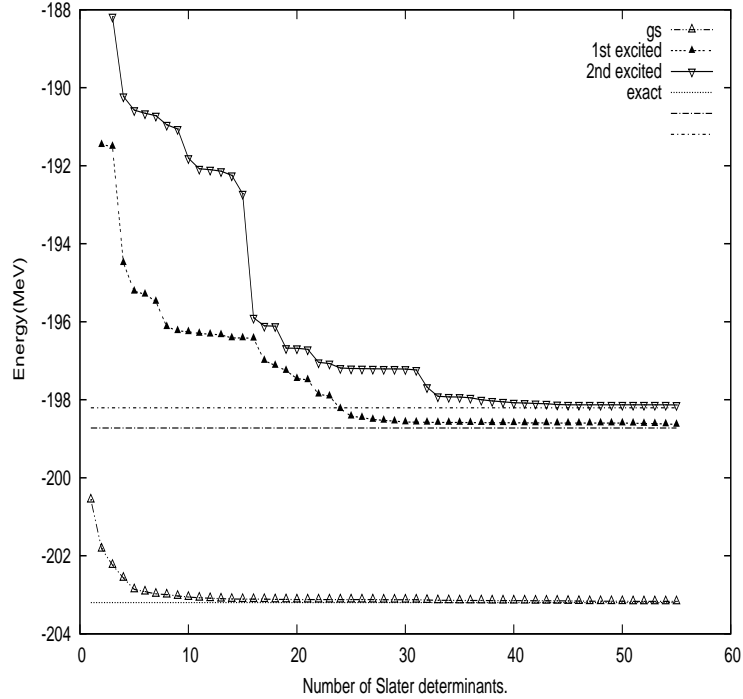}
\caption{Energies of $0^+$ states as a function of the number of Slater determinants.}
\end{figure}
\renewcommand{\baselinestretch}{2}
     For this set of states we considered also a few more calculations with all non-zero $w_1,w_2,w_3$.
     Initially the ground-state is strongly overweighted, then as the number of Slater determinants is increased
     we allow for more balanced values and at the end $w_1=3,w_2=2.5,w_3=1$. For a set of $50$ Slater determinants
     we obtained $E(0^+_1)=-203.167$MeV, $E(0^+_2)=-198.637$MeV, $E(0^+_3)=-198.130$MeV, in line with the previous results.
\par  
     For the $2^+$ states we evaluated the first $3$ states, as before. We have used $50$ Slater determinants and
     used non-zero values for the weights. Our results are $E(2^+_1)=-200.122$MeV, $E(2^+_2)=-198.064$MeV and
     $E(2^+_3)=-196.725$MeV. The energy of the $2^+_1$ state is about $70$KeV higher than the shell model value.
     The shell model values for the other states are not available. Although, there is still some room for a further decrease
     in the energies, the excitation energies seem to be stable enough as a function of the number of Slater
     determinants, whithin an uncertainty of few tens of KeV.
\par
     In conclusion, we have presented a simple method to evaluate the non-yrast eigenstates with minimal, if any,
     modifications to the ground-state computer programs. 

\vfill
\eject

\begin{thebibliography}
\bigskip
\bibitem{1}
       G.Puddu. J. Phys. G: Nucl. Part. Phys. {\bf 32},321 (2006).
\bibitem{2}
       K.W.Schmid, Zheng Ren-Ron, F.Grummer and Amand Faessler.\\
          Nucl. Phys.  {\bf A 949},63(1989).
\bibitem{3}
       N.Shimizu,Y.Utsuno, T.Mizusaki, T.Otsuka, T.Abe and M.Honma.\\
        Phys. Rev. {\bf C 82},061305 (2010).
\bibitem{4}
       N.Shimizu. Private communications.
\bibitem{5}
       G.Puddu. Eur. Phys. J. {\bf A 31}, 163(2007).
\bibitem{6}
      K.W.Schmid and F.Grummer. Rep. Progr. Phys. {\bf 50},731(1987).
\bibitem{7}
      G.Puddu. Eur. Phys. J {\bf A 42},281(2009).
\bibitem{8}
     W. Lederman ed. Handbook of Applicable Mathematics. Vol. III,\\
      Numerical Methods, chapter 11. John Wiley and Sons, New York 1981.
\bibitem{9}
      W.A.Richter, M.G. van der Merwe, R.E.Julies and B.A.Brown.\\
      Nucl. Phys. {\bf A 523},325(1991).
\end{thebibliography}
\end{document}